# DIGITAL LONG FOCAL LENGTH LENSLET ARRAY USING SPATIAL LIGHT MODULATOR

Akondi Vyas[1,2], M B Roopashree[1], B R Prasad[1]

[1]Indian Institute of Astrophysics, 2nd Block, Koramangala, Bangalore
[2]Indian Institute of Science, Bangalore
vyas@iiap.res.in, roopashree@iiap.res.in, brp@iiap.res.in

**Abstract:** Under a thin lens and paraxial approximation, the phase transformation function of a lens was simulated on a Liquid Crystal (LC) based Spatial Light Modulator (SLM). The properties of an array of such lenses simulated on transmitting type and reflecting type SLMs were investigated and the limits of its operation in wavefront sensing applications are discussed.

## 1. INTRODUCTION

A Shack Hartmann Sensor (SHS) is a crucial and most commonly used wavefront sensing element in adaptive optics technology, widely used for compensating dynamic turbulence effects in real time imaging systems [1]. SHS is an array of lenses that reads the wavefront shape in terms of the shift in the spots of the sensor at the focal plane of the lenses. Liquid Crystal (LC) based Spatial Light Modulator (SLM) can be used to modulate the phase of the incident beam of light [2]. These devices can be used to imitate the behavior of a SHS by projecting Diffractive Optical Lenses (DOLs) on them [3]. The advantage of SLM based SHS is that its properties can be dynamically controlled. The disadvantage of the DOL based SHS is their focusing inefficiency [4]. The thickness function of a lens governs its phase transformation function. It is possible to realize a lens on an SLM by addressing suitable grayscale values that replicate the thickness function.

In this paper, we present the theoretical design and experimental results of a digital lens based SHS. It is essential to determine the performance metrics of the lenslet array to determine optimum conditions [5]. The major sources of error include temporal instabilities, calibration and alignment errors and centroid estimation errors due to low signal to noise ratio, background noise and detector noise [6]. Experiments were performed to measure the focal length of the lenslet array, beam profile and the extent of temporal stability. The sensitivity of the device to imperfect positioning of the detector with respect to the incoming beam was quantified.

## 2. DESIGN OF LENSLET ARRAY

Under a thin lens approximation, the thickness function of a lens, $\Delta(x, y)$ can be written as [7],

$$\Delta(x,y) = \Delta_0 - R_1\left(1 - \sqrt{1 - \frac{x^2+y^2}{R_1^2}}\right) + R_2\left(1 - \sqrt{1 - \frac{x^2+y^2}{R_2^2}}\right) \quad (1)$$

where, $\Delta_0$ is the maximum thickness of the lens and $R_1$, $R_2$ are the radii of curvature for surfaces of the lens on the left and right respectively. Further approximating equation (1) by considering only paraxial rays, the thickness function becomes [7],

$$\Delta(x,y) = \Delta_0 - \frac{x^2+y^2}{2}\left(\frac{1}{R_1} - \frac{1}{R_2}\right) \quad (2)$$

Considering a double convex lens,
$$R_1 = -R_2 = R \quad (3)$$
From equations (2) and (3), we obtain,

$$\Delta(x,y) = \Delta_0 - \frac{(x^2+y^2)}{R} \quad (4)$$

It is possible to make a converging lens out of the SLM by addressing gray scale command values that follow a phase function of the form shown in equation (4). Lens-Maker's formula for a thin convex lens gives the focal length (f) of the lens,

$$\frac{1}{f} = (n_{eff} - 1)\left(\frac{2}{R}\right) \quad (5)$$

where $n_{eff}$ is the effective refractive index of the LC-SLM display. $\Delta_0$ is equivalent to $\Delta_{Max}$, the maximum phase that can be obtained using the SLM. Hence equation (4) can also be written as:

$$\Delta(x,y) = \Delta_{Max} - \frac{(x^2+y^2)}{R} \quad (6)$$

Fixing phase reference above zero,

$$\Delta_{Max} - \frac{x^2+y^2}{R} > 0 \quad (7)$$

hence, $\quad R > \frac{x^2+y^2}{\Delta_{Max}} \quad \forall \quad (x,y) \quad (8)$

Consider a lens diameter of 'N' pixels on the SLM and if 'p' is the SLM pixel pitch, then,

$$R > \frac{(Np/2)^2}{\Delta_{Max}} = \frac{N^2 p^2}{4\Delta_{Max}} = R_{optimum} \quad (9)$$

The choice of R=$R_{optimum}$ will allow the use of full range of SLM phase and hence equation (4) becomes,

$$\Delta(x,y) = \Delta_{Max}\left[1 - \frac{4(x^2 + y^2)}{N^2 p^2}\right] \quad (10)$$

The effective maximum thickness of the display can be calculated as

$$d_{eff} = \frac{\Delta_{Max}}{n_{eff}} \quad (11)$$

Combining equations (5), (9) and (11)

$$N = C\sqrt{f} \quad (12)$$

$$\text{where,} \quad C = \left\{\frac{8\Delta_{Max}}{p^2}\left(\frac{\Delta_{Max}}{d_{eff}} - 1\right)\right\}^{\frac{1}{2}} \quad (13)$$

Since 'C' can be measured experimentally by measuring 'f' at different lens diameters, $d_{eff}$ and $n_{eff}$ can be estimated. This method can be very simple and useful technique to estimate the effective refractive index of LC-SLMs.

Lenses were simulated on transmitting type (T-SLM, Holoeye LC 2002) and reflecting type (R-SLM, Holoeye LC-R 720) SLMs. An inversion methodology was adopted before applying the phase generated using equation (10) to overcome the nonlinear behavior of SLMs [8].

At 633nm, for T-SLM and R-SLM, the phase values are applied such that,

$$\Delta_T(x,y) = 0.27\mu m\left[1 - \frac{4(x^2 + y^2)}{N^2 p_T^2}\right] \quad (14)$$

$$\Delta_R(x,y) = 0.34\mu m\left[1 - \frac{4(x^2 + y^2)}{N^2 p_R^2}\right] \quad (15)$$

$p_T$ and $p_R$ represent the pixel pitch of T-SLM and R-SLM respectively.

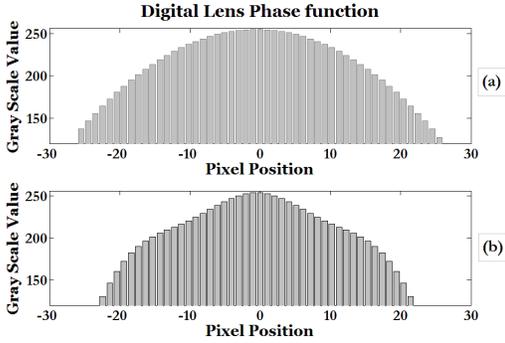

Fig. 1. (a) Calculated phase transformation of lens (b) applied grayscale command values for T-SLM

To generate a circular aperture, $\Delta(x,y)$ was calculated at all discrete 'x' and 'y' such that,

$$\sqrt{x^2 + y^2} < \frac{Np}{2} \quad (16)$$

A comparison of the phase transformation function (corresponding grayscales) computed using equation (15) and the grayscale command values (after nonlinear inversion) to be applied to get the desired phase is shown in Fig. 1 for the T-SLM. The obtained spot pattern by applying a phase function in equation (14) on T-SLM is shown in Fig. 2.

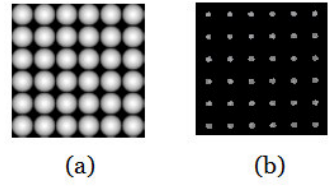

Fig. 2. (a) Pixel values addressed on T-SLM, N=26. (b) Spot pattern recorded near focal plane of lenses projected on T-SLM

### 3. CALIBRATION EXPERIMENTS

The lenslet calibration analysis includes beam profile measurement, testing temporal stability of the focal spots, the centroiding accuracy and sensitivity of the device to alignment errors [5].

### 3.1 Focal Length measurement

Lenses of different 'N' were simulated by using equations (14) and (15) for T-SLM and R-SLM. To measure the focal length of the lenses, a simple experimental setup was used as shown in Fig. 3.

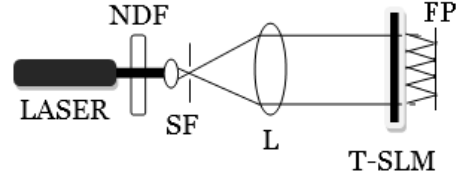

Fig. 3. Schematic of the experimental setup to measure focal length for T-SLM

Melles Griot He-Ne laser (633nm) was used as a source of light. A neutral density filter (NDF) was used to control the beam intensity. The beam was cleaned using a compact five axis Newport spatial filter (SF) setup. A 12.5cm triplet lens, 'L' was used to collimate the beam. T-SLM (Lens phase transformation function imposed) was placed in the path of the parallel beam and the focal plane was captured on a CCD (Pulnix TM 1325CL) by mounting it on a translation stage. The results of measured focal length for T-SLM are tabulated in Table 1.

Table 1. Measured Focal length for T-SLM

| Pixels used for a single aperture | Measured Focal length (cm) | Theoretical Focal length (cm) |
|---|---|---|
| 26 | 30 | 27.35 |
| 39 | 62 | 61.53 |
| 48 | 90 | 93.21 |
| 52 | 112 | 109.39 |

## 3.2 Temporal Stability

The stability of the DOL based SHS on SLM has been reported recently [9]. Using a similar configuration shown in Fig. 3, we monitored the spots for 10 min by recording the spot pattern images in steps of 1sec on the CCD. Iteratively weighted centroiding technique (total number of iterations used = 3) was used to locate the center of the spot [10]. The position of the spot center varies in x and y directions with a standard deviation, σ of 0.28. The graph illustrating the variability of the spot position is shown in Fig. 4. In order to overcome this error in the measurements, it is important to consider the fact that the minimum detectable shift in the spot must be $3\sigma_{max} \sim 0.84$ pixels (where, $\sigma_{max} = \max\{\sigma_x, \sigma_y\}$). This limits the minimum readable phase difference by the wavefront sensor. Hence, the minimum detectable phase difference with a shift of 0.84 pixels on the detector for the case of 26 pixel lens on T-SLM with $f \sim 30$cm is $> \lambda/10$. This wavelength dependent quantity varies with aperture size, focal length and the minimum shift in the spot above the error threshold (determined by the spot stability). The effect of power fluctuations of the laser and the detector noise were not removed in the stability analysis.

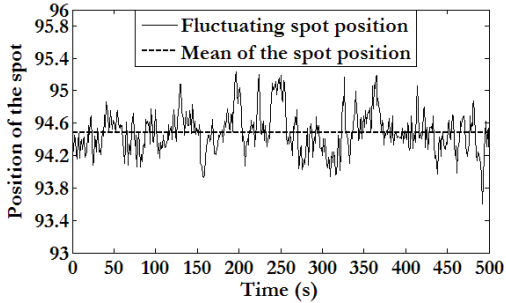

Fig. 4. Temporal Stability of the SHS spot pattern

### 3.3 Beam profile measurement

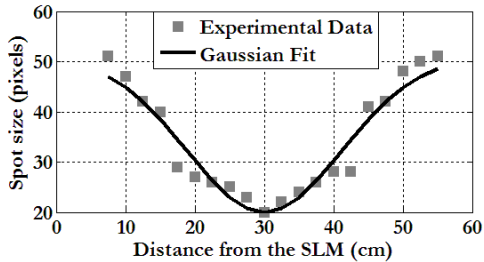

Fig. 5. Beam profile measurement for T-SLM with 26 pixel lens

A precise measurement of the beam profile was made by measuring the spot size at different distances from the lens. In the case of DOLs, the outgoing beam is non Gaussian and the focusing depends on the interfering spherical waves. In our case, the beam is more close to a Gaussian and follows the structure shown in Fig. 5.

### 3.4 Sensitivity of detector position

An adaptive optics system is highly sensitive to the position of the sensor. Experimental setup for testing the sensitivity of rotation of the lenslet array is similar to the one shown in Fig. 3 with a difference that an aberrator was placed between the lens L and T-SLM. A compact disc case was used as an aberrator since it can replicate a Kolmogorov phase screen [11, 12]. The SLM panel was rotated in the plane parallel to the horizontal in steps of $0.02^0$. Using a 12×12 lenslet array, 288 slopes (144 each in x, y directions) were measured at each step. The number of slopes that change below a threshold (0.1 pixels) for each step of the angle of rotation was counted. This number quantifies the sensitivity of the sensor to small rotations; larger the number, better the wavefront reconstruction accuracy.

The detector has to be placed near the focal plane of the SHS. To characterize the sensitivity of the position of the CCD, CCD was translated about the focal plane and the shape of the reconstructed wavefront was compared with the one at focus using correlation coefficient.

As can be seen from Fig. 5, the lens simulated on the SLM has a large confocal parameter (> 16cm) and hence by maintaining a proper reference spot pattern (recorded by allowing a plane parallel beam to be incident on SHS), it is possible to eliminate the errors due to the misalignment of the CCD and the SLM.

### 3.5 Calibration of shift in spot

To produce controlled known phase differences, a well calibrated reflecting type SLM was used [13]. The schematic for the measurement of the shift in the spots of the SHS is shown in Fig. 6.

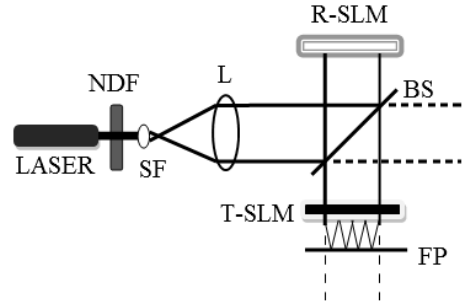

Fig. 6. Schematic for SHS shift calibration

The collimated beam is allowed to pass through a beam splitter and incident on the R-SLM. The plane wave incident on the R-SLM reflects a wavefront shaped in accordance with the addressed phase distribution. This wave-front is then sensed by the T-SLM based SHS. Phase differences starting at 0.1λ to

0.55λ were applied in steps of 0.05λ across the diameter of a single aperture on T-SLM. The resultant translation of the focal spot in x and y directions is recorded and the spot position is identified after applying intensity weighted centroiding [14]. Fig. 7 shows the drift in the spot position with changing phase difference across a single subaperture. The imperfections in the shift are due to the small amplitudes of phase differences applied. The observed shift in the spot (~ 4 pixels on CCD for a phase difference of 0.45λ) nearly matches with the theoretically expected value (exactly 4.24 pixels) for a converging lens of same focal length (f = 30 cm; N=100 on T-SLM).

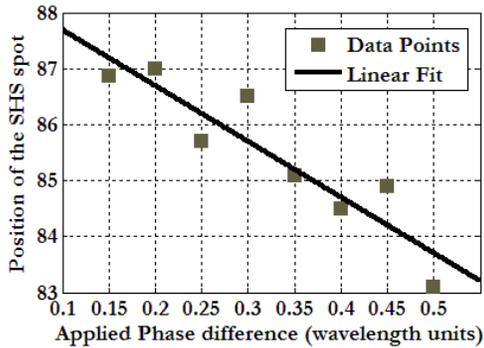

Fig. 7. Shift in the spot calibration of SHS

## 4. CONCLUSION

It was shown that a digital Shack Hartmann sensor can be simulated on a liquid crystal based SLM. The effective refractive index of the LC-SLM material was measured to be ~ 2 for both the SLMs. The lens followed a Gaussian beam profile. The centroid position of the spot had a standard deviation of 6μm under undisturbed conditions. It was observed that positioning of the detector exactly at the focal plane of the lens is not critical. This is because of a very long focal length and confocal parameter of the lenslets. The shift in the spots was linear with respect to the applied phase difference.